\documentstyle[prl,aps,twocolumn,epsfig,floats]{revtex}

\newcommand{\sca}{scaling}

\begin{document}
\wideabs{
%\twocolumn[
%\hsize\textwidth\columnwidth\hsize\csname@twocolumnfalse\endcsname
\draft

\title{Precursory dynamics in threshold systems}
\author{J.S. S\'a Martins $^{\dag}$, J.B. Rundle $^{\dag}$, 
M. Anghel $^{\ddag}$, and W. Klein $^{\ddag\pm}$}
\address{$\dag$ Colorado Center for Chaos and Complexity/CIRES and
  Department of Physics, CB 216, University of Colorado, Boulder,
  Colorado 80309}
\address{$\ddag$ Center for Nonlinear Studies, LANL, Los Alamos, New
  Mexico 87545}
\address{${}^{\pm}$Permanent address: Physics Department and Center 
for Computational Science, Boston University,  Boston, MA 02215}
\date{\today}
\maketitle

%\widetext
\begin{abstract}
A precursory dynamics, motivated by the analysis of recent experiments on 
solid-on-solid friction, is introduced in a continuous cellular automaton 
that mimics the physics of earthquake source processes. 
The resulting system of equations for the interevent cycle can be decoupled 
and yields an analytical solution in the mean-field limit, exhibiting a  
smoothing effect of the dynamics on the stress field. 
Simulation results show the resulting departure from scaling at the 
large-event end of the frequency distribution, and support claims that the 
field leakage may parameterize the superposition of {\sca} and characteristic 
regimes observed in real earthquake faults.
\end{abstract}
%\maketitle
}
%]

%\begin{twocolumn}
%\begin{multicols}{2}
%\narrowtext

There are many examples in nature of systems whose internal dynamics have as 
an essential feature the separation of time scales into a long-term loading 
process and a short-term discharge. In the former, some external source 
slowly drives the variables that control the dynamics until these fields 
reach some threshold value. When this happens, the fast dynamics of the 
discharge takes over and generates a sequence of internal rearrangements, 
called an avalanche in this context, which eventually removes the excess 
through internal dissipation, or by flow across the system's borders. These 
dynamical threshold systems have been the object of much recent interest, 
due mainly to the fact that they can be recast into a discrete time-evolution 
form. The resulting discrete dynamics transforms the fast time scale discharge 
cycle into a sequence of well-defined time steps, in each of which the
dynamical fields undergo a stochastic transformation that depends on
their configuration in the previous step as well as annealed noise. As
such, and since the fields are in general real valued, we will call
these representations stochastic continuous cellular automata (SCCA). 

Physical systems that have been modeled and studied along these lines are 
abundant in the literature, ranging from the firing behavior of neural 
networks \cite{neural} to the dynamics of domain walls in magnets 
\cite{sethna}, the motion of vortex lines in type II superconductors 
\cite{suc}, the depinning transition in the growth of interfaces in
random media \cite{paczuski}, the dynamics of the energy release in
solar flares \cite{scott}, and to the source processes responsible for 
earthquakes.

The existence of an interevent dynamical cycle in the latter is
suggested by recent laboratory experiments addressing issues of 
solid-on-solid friction. A stable slip, with a slow velocity that
increases with the stress level, is observed prior to failure, leading
to a partial release of the accumulated stress \cite{lab,tullis}. This
stress leakage mechanism is analogous to a temperature-dependent
viscosity that has been observed in laboratory for the creeping of 
crystalline rocks, and can be modeled by the equation

\begin{equation}
\frac{d s(t)}{dt} = \alpha \frac{\sigma(t) - \sigma_R}{K}
\label{leak}
\end{equation}
where $s(t)$ and $\sigma(t)$ are the displacement and the stress at time $t$, 
$\sigma_R$ is some residual stress value to which the system decays after 
$\sigma(t)$ reaches a failure threshold $\sigma_F$, $K$ is the elastic 
stiffness and $\alpha$ measures the intensity of this stress leakage effect. 
This parameter, with dimensions of inverse time, will in general be dependent 
on both the stress level and the temperature. An analogous leakage mechanism 
has also been suggested in the context of integrate-and-fire neural 
networks \cite{neural}, and it is likely that the results reported
here will also hold in that context.

As a summary, we will show that the introduction of a stress leakage process 
as an interevent dynamics in a SCCA model for earthquake faults changes in a 
dramatic way the space-time patterns it generates. In particular, the 
$\alpha$ value for a fault may determine its overall behavior as of a 
scale-invariant type or a nucleation type, with a mixed composition in
between, reproducing features observed in real faults \cite{wesnousky}. The 
importance of this new parameter in earthquake source models has in
fact been recently evaluated. Its tuning to match the characteristics
of each segment in a complex computer representation of the fault
network of southern California allowed the generation of space-time
patterns of rupture of unprecedented realism \cite{paul}.

We will consider here the Rundle-Jackson-Brown (RJB) SCCA model for an 
earthquake fault \cite{rjb}, in its uniform long-range interaction, 
mean-field version \cite{cise}. 
Extensive work has recently focused a near-mean-field version of this model, 
where the interaction range has a cutoff and the model can be mapped onto an 
Ising-like Langevin equation \cite{agu}. Its dynamical variables are two 
continuous real-valued fields, slip $s_i(t)$ and stress $\sigma_i(t)$, defined 
on the sites $i$ of a lattice. A constitutive equation couples these fields,

\begin{equation}
\sigma_i(t) = \sum_j T_{ij} s_j(t) + K_Lvt
\label{const}
\end{equation}

\noindent
where $T_{ij}$ is the interaction matrix, or stress Green's function, 
$T_{ii} = - K$, $K = K_L + K_C$, $K_C = \sum_{j \neq i} T_{ij}$, $K_L$ 
parametrizes the loading interaction, and 
$v$ fixes a relation between the short and long time scales of the physics 
of the model. Equation (\ref{const}) 
specifies the stress at each site as long as 
max$[\sigma_i(t)] < \sigma_F$, 
which defines the stress threshold $\sigma_F$, taken as uniform over the 
lattice. The interevent, or loading, dynamics of the standard RJB model is 
thus very simple; the slip field remains static while the stress field 
undergoes linear growth. As soon as the stress at one site reaches the
threshold, a fast stochastic relaxation dynamics takes over. The slip
at a site that fails is discontinuously reset to a new value that
leads the stress to assume its residual value, usually with some
noise, introduced to represent the disorder in the rheology. The stress
drop is redistributed among the interacting neighbors, with an
intrinsic dissipation measured by the factor $\delta = K_L/K$:

\begin{equation}
\Delta \sigma_j = \frac{T_{ji}}{K} |\Delta \sigma_i|,
\end{equation}
where $i$ is the site that failed. This increase may cause other sites to 
fail as well, and the process continues until all sites have stress below 
failure. This cascade of failures, or avalanche, is the model's equivalent 
for an earthquake.

The relaxation dynamics of the model can be cast into a single field 
formulation that is specially convenient for computer simulations. 
In units of the short time scale,

\begin{eqnarray}
\nonumber
\sigma_i(t+1) &=& \sigma_i(t) + \sum_j \left\{ T_{ij} 
\frac{\sigma_j(t) - \sigma_R}{K} \Theta \left( \sigma_j(t) - \sigma_F \right) 
\right\}\\ 
&+& \xi_i(t)
\label{relax}
\end{eqnarray}
where $\xi_i(t)$ is the noise term. We direct the reader to
Ref. \cite{cise} for a complete and pedagogical discussion of this
relaxation dynamics.

We report in this paper results obtained for the 
introduction of stress leakage, as modeled by Eq. (\ref{leak}), as an 
interevent dynamics of our SCCA model. The equations for these
intervent dynamics are obtained by taking the time derivative of
Eq. (\ref{const}), substituting $ds_i(t)/dt$ from
Eq. (\ref{leak}), and choosing, with no loss of 
generality, $\sigma_R = 0$, to get,

\begin{equation}
\frac{d\sigma_{i}(t)}{dt} = \frac{\alpha}{K} \sum_{j} T_{ij} \sigma_{j}(t) 
+ K_L v,
\label{eqset}
\end{equation}
a set of $N$ coupled equations for the stress field. Without
introducing any particular form for the $T_{ij}$, these equations
can be combined to derive the time evolution of the average stress 
for short times
$\bar \sigma(t) = \sum_i \sigma_i(t) / N$ :

\begin{equation}
\frac{d \bar \sigma(t)}{dt} = - \alpha \frac{K_L}{K} \bar \sigma(t) + K_L v,
\end{equation}
with the solution, cast in dimensionless form by defining  
$\tilde t = \sigma_F/K_Lv, \eta = \sigma/\sigma_F, 
\tau = t/\tilde t$, and $\phi = \alpha \tilde t$,

\begin{equation}
\bar \eta(\tau) = \left(\bar \eta(0) - \frac{1}{\delta \phi}\right)\, 
exp({- \delta \phi \tau}) + \frac{1}{\delta \phi},
\end{equation}
which is an increasing (decreasing) function of dimensionless time if
$\bar \eta(0) < (>) 1/\delta \phi$

Equations \ref{relax} and \ref{eqset} above define the generic RJB
model with stress leakage for the interevent dynamics. 
The current focus of research is on the long-range versions, where the 
interaction matrix $T_{ij}$ is nonzero for a substantial fraction of all 
pairs $(ij)$. Ideal elasticity indicates that this matrix should involve a 
dependence on the distance as $1/r^3$, as expected from the results of 
simple laboratory experiments. Nevertheless, it has been shown that the upper 
critical dimension for this interaction is $d_u=2$ \cite{fisher} and we can 
recover the long wavelength physics of this model by a simpler mean-field 
formulation. We will thus focus on the mean-field RJB model, with a uniform 
interaction matrix 
$T_{ij} = K_c/(N - 1)$, where $N = 1/\Delta$ is the number of 
sites in the lattice. In this case, the set of equations in (\ref{eqset}) can 
be decoupled and yield a solution which reads, in dimensionless
variables, 

\begin{eqnarray}
\nonumber
\eta_i(\tau) &=& \left[ \eta_i(0) - \bar \eta(0) \right] exp \left\{ 
- \left[ \frac{1 - \delta \Delta}{1 - \Delta} \right] \phi \tau \right\}\\ 
&+& \left(\bar \eta(0) - \frac{1}{\delta \phi}\right) exp(-\delta \phi \tau) 
+ \frac{1}{\delta \phi}
\label{stressevo}
\end{eqnarray}

From
%\begin{equation}
\[
\eta_i(\tau) - \eta_j(\tau) = [\eta_i(0) - \eta_j(0)] exp \left\{ 
- \left[ \frac{1 - \delta \Delta}{1 - \Delta} \right] \phi \tau
\right\}
\]
%\end{equation}
we can see that this time evolution of the stress field is order-preserving, 
i.e.
$\eta_i(0) > \eta_j(0) \Rightarrow \eta_i(t) > \eta_j(t)$, 
allowing for the determination of the initiator of the next event by finding 
the site with maximum stress at the completion of the previous event. The 
important effect of this leakage stress dynamics in the pattern of failures of 
the system comes from the reduction it causes in the statistical spread of the 
stress field with time. A simple measure of this smoothing is obtained through 
the time evolution of the variance of the stress field

\begin{equation}
var\left[\eta(t)\right] = var\left[\eta(0)\right] exp \left\{
- \left[ \frac{1 - \delta \Delta}{1 - \Delta} \right] \phi \tau \right\}
\end{equation}
Because of this exponential smoothing of the stress field, the probability 
of a site to reach failure after receiving a transfer from a failing neighbor 
increases, and is an increasing function of the time to failure. As a 
consequence, the branching ratio, defined as the average number of failures 
caused by each failing site, also increases. The system is more likely to 
undergo larger avalanches, which may even be system wide when the 
time to failure is large enough.

Because this interevent dynamics preserves order, as already mentioned, its 
introduction in a SCCA is rather straightforward. The site with stress nearest 
to failure $\eta_{max}$ after an event is the initiator of the next, and 
solving for $\tau$ in Eq. (\ref{stressevo}) for this site determines the 
time-to-failure $\tau_F$, that will be used again to 
update the stress field over the lattice prior to the event. To avoid the 
time-consuming solving of transcendental equations at each time step in the 
simulation, the time-evolution equations may be linearized, as long as
max $\{ \delta \phi \tau,[(1 - \delta \Delta)/(1 - \Delta)] \phi \tau \} 
<< 1$, which is true for our parameters, to read

%\begin{equation}
\[
\eta_i(\tau) = \eta_i(0) + \left\{ 1 + \left[ \frac{1 - \delta}{1 - \Delta} 
\right] \phi \bar \eta(0) - \left[ \frac{1 - \delta \Delta}{1 - \Delta} 
\right] \phi \eta_i(0) \right\} \tau\\ 
\label{stressevolin}
\]
%\end{equation}
and the solution for the time-to-failure is

\begin{equation}
\tau_F = \frac{1 - \eta_{max}}{1 + \left[ \frac{1 - \delta}{1 - \Delta} 
\right] \phi \bar \eta(0) - \left[ \frac{1 - \delta \Delta}{1 - \Delta} 
\right] \phi \eta_{max}}
\end{equation}

\begin{figure}[htb]
\centerline{\psfig{figure=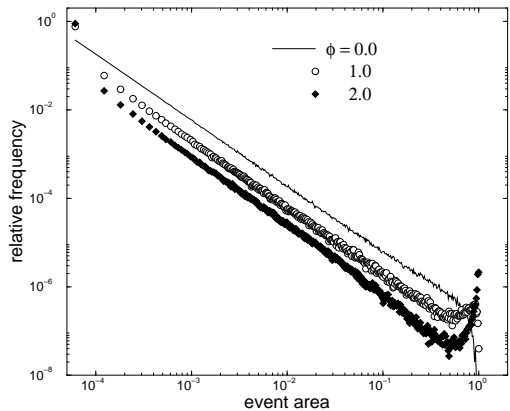, width=5.5cm, angle=270}}
\caption{The frequency distribution is shown for dimensionless leakage
  factors $\phi = 1.0$ and $2.0$; the plot with no leakage is shown
  for a comparison. The simulations were performed on a $128 \times
  128$ lattice, with a dissipation factor $\delta = 0.01$ and a noise
  amplitude of $0.5$. Data for this and all subsequent plots was
  collected from $3 000 000$ events, after a transient of the same
  order, and was logarithmically binned.}
\label{scaling}
\end{figure}

Figure (\ref{scaling}) shows a log-log plot of the frequency distribution of 
events as a function of their size. The effect of the stress leakage dynamics 
shows up clearly in the excess over scaling obtained for large events as 
$\phi$ is increased from $0$, together with a depletion of the distribution 
in the intermediate size range. The slope of the scaling part of the plot 
also gradually increases, starting from the mean-field value $\tau = 1.5$. 
The smoothing effect of the leakage dynamics, together with the
resulting larger stress average that it causes in the field as a
whole, results in a higher probability for large events to grow, 
eventually causing total rupture of the fault \cite{eric}.

\begin{figure}[htb]
\centerline{\psfig{figure=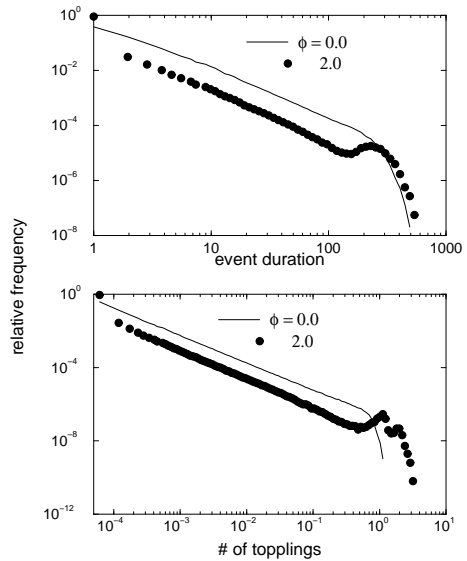, width=7.5cm, angle=270}}
\caption{The frequency distribution for the number of topplings and
  for the dimensionless event duration are shown for effective leakage
  factor $\phi = 2.0$; the plot with no leakage is shown for a comparison.}
\label{duration}
\end{figure}

This smoothing effect is reflected also in the distribution for the number 
of topplings, shown in the lower plot on Fig. \ref{duration}. Here, a
counter is
updated each time a site fails, even if it had failed before in the
same avalanche. This number reflects more closely the model's
equivalent for the moment release in an event. The fraction of
multiple failures vanishes in the exact mean-field limit, and the
number of topplings become equivalent to the event area. This is no
longer true for the model with stress leakage. The single maximum at
the upper end of Fig. \ref{scaling} corresponds, in fact, to very 
different moment 
releases, as shown in Fig. \ref{duration}. An intriguing feature of the 
latter is the double bump at the large end, located close to integer 
multiples of the system size. They suggest that total rupture of the system 
is more likely, in this region, than one would naively expect.

The upper part of this plot shows the distribution of event durations, 
defined in the model as the number of updates in the fast time scale that are 
required for relaxation. Again, the model with leakage reflects an excess 
of longer events over scaling: events with some range of large durations are 
more frequent, and the plot shows a local maximum close to its high
end.

\begin{figure}[htb]
\centerline{\psfig{figure=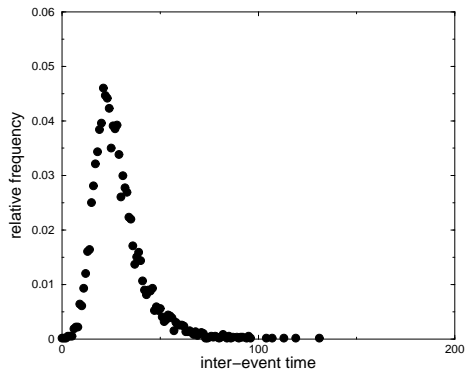, width=5cm, angle=270}}
\caption{The frequency distribution for dimensionless interevent times
  between large events that rupture a fraction of 0.9 of the total sites is
  shown for the dimensionless leakage factor $\phi = 2.0$. The plot shows the 
  establishment of a characteristic event regime, superimposed on a
  complex background.}
\label{char}
\end{figure}

The SCCA models with no leakage dynamics show no signs of a 
characteristic-event regime. The power spectrum of the distribution of 
interevent times for large size events is white. The inclusion of leakage 
dynamics, however, changes this aspect radically, as shown in Fig. \ref{char}. 
The distribution of interevent times has a pronounced maximum, corresponding 
to a characteristic period between large rupture events. This feature is 
tuned by the leakage parameter, and will be more dominant as it
increases.

\begin{figure}[htb]
\centerline{\psfig{figure=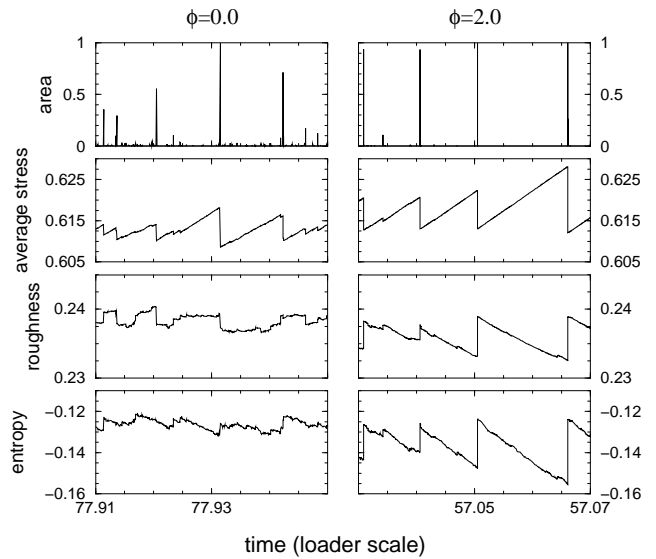, width=7.5cm, angle=270}}
\caption{The plot compares the time evolution of the dimensionless
  average stress, roughness, and configuration entropy for the model
  with no leakage ($\phi=0$) and with leakage ($\phi=2.0$). The upper
  plot shows the area of events in the time window selected, as a fraction of
  the lattice area. The model with leakage shows a much more regular
  dynamics and allows a larger buildup of the average stress, together
  with smaller roughness and entropy, which leads to a characteristic
  event regime.}
\label{window}
\end{figure}

The effects of the leakage dynamics on the statistical properties of the 
model's stress field are made more explicit in Fig. \ref{window}. This plot 
shows the time evolution, in the slow time scale, of its average value and 
roughness, together with the configuration entropy. This last quantity is a 
measure of the degree of ordering of the stress distribution \cite{dahmen}. 
Notice that the average stress is an increasing function of the time between 
large events, both for the model without and with leakage, but that in the 
latter the overall average is higher. As opposed to what was seen in a SCCA 
model with no leakage but with varying dissipation and weakening of failed 
sites \cite{dahmen}, we do not recognize a mode switching dynamics as present 
in our case. It remains to be seen what features would result from a 
combination of these dynamics.

We presented in this paper a SCCA model with aseismic
creep superimposed on a seismic threshold dynamics, as recent
laboratory experiments have shown to exist on solid-on-solid
friction. In a mean-field approximation, with infinite range of uniform 
interactions, the equations for the interevent stress time evolution
can be solved. Computer simulations of the resulting SCCA model,
combining the two dynamics, show a superposition of complex time 
and space patterns with a more regular occurrence of characteristic events, 
with rupture of the entire fault. 

%\smallskip
%\centerline{\bf Acknowledgments}

Research by J.S.S.M. was supported by CIRES, University 
of Colorado at Boulder, J.B.R. was supported by DOE (Grant No. 
DE-FG03-95ER14499), M.A. by DOE (Grant No. W-7405-ENG-36), and W.K. by
DOE (Grant No. DE-FG02-95ER14498).

%\smallskip

%\end{twocolumn}
%\end{multicols}
\end{document}